# Crystalline structure and XMCD studies of $Co_{40}Fe_{40}B_{20}$ grown on $Bi_2Te_3$, BiTeI and $Bi_2Se_3$

*A. K. Kaveev*[1], N. S. Sokolov[1], S. M. Suturin[1], N. S. Zhiltsov[1], V. A. Golyashov[2], O. E. Tereshchenko[2], I. P. Prosvirin[3], K. A. Kokh[4] and M. Sawada[5]

[1]Ioffe Institute, St.- Petersburg, Russia
[2]Rzhanov Institute of Semiconductor Physics, Novosibirsk, Russia
[3]Boreskov Institute of Catalysis, Novosibirsk, Russian Federation
[4]Novosibirsk State University, Novosibirsk, Russia
[5]University of Hiroshima, Hiroshima, Japan

*Abstract.* Epitaxial films of $Co_{40}Fe_{40}B_{20}$ (further - CoFeB) were grown on $Bi_2Te_3$(001) and $Bi_2Se_3$(001) substrates by laser molecular beam epitaxy (LMBE) technique at 200-400°C. Bcc-type crystalline structure of CoFeB with (111) plane parallel to (001) plane of $Bi_2Te_3$ was observed, in contrast to polycrystalline CoFeB film formed on $Bi_2Se_3$(001) at RT using high-temperature seeding layer. Therefore, structurally ordered ferromagnetic thin films were obtained on the topological insulator surface for the first time. Using high energy electron diffraction (RHEED) 3D reciprocal space mapping, epitaxial relations of main crystallographic axes for the CoFeB/ $Bi_2Te_3$ heterostructure were revealed. MOKE and AFM measurements showed the isotropic azimuthal in-plane behavior of magnetization vector in CoFeB/ $Bi_2Te_3$, in contrast to 2nd order magnetic anisotropy seen in CoFeB/$Bi_2Se_3$. XPS measurements showed more stable behavior of CoFeB grown on $Bi_2Te_3$ to the oxidation, in compare to CoFeB grown on $Bi_2Se_3$. XAS and XMCD measurements of both concerned nanostructures allowed calculation of spin and orbital magnetic moments for Co and Fe. Additionally, crystalline structure and XMCD response of the CoFeB/BiTeI and $Co_{55}Fe_{45}$/BiTeI systems were studied, epitaxial relations of main crystallographic axes were found, and spin and orbital magnetic moments were calculated.

## Introduction

Last years pronounced interest in spintronics area is focused on the novel class of the materials – topological insulators (TI). These materials possess an insulating properties in a bulk, at that their surface is conductive, because of the strong spin-orbital interaction in TI. This interaction results in spin splitting of the surface states forming Dirac cone near Γ point [1]. Studies of the ferromagnetic (FM, or ferromagnetic insulator) - TI interface are attractive due to the magnetic proximity effect, which acts as symmetry breaking factor relative to the topological states time inversion. This effect appears in FM exchange field influence on the TI surface states. The inverse effect is of interest also: the spin-polarized current flow through the topological states of TI may cause different types of the magnetic order in FMs. An idea of the mutual tuning between the TI interfacial states and FM magnetization is perspective for the development of different spintronic devices based on the magneto-resistive systems, FM spin transistors (SpinFETs) and spin batteries [2, 3].

To create SpinFET, it is necessary to inject spin-polarized carriers from the FM contact into the TI topological states and to detect spin current with the second FM contact with another value of a coercivity. In present paper we have carried out a step to obtain high quality FM/TI interface which provides low decrease scattering of the spin-polarized carriers. Obtaining the structurally ordered FM on TI will also decrease the scattering. For this purpose we have grown $Co_{40}Fe_{40}B_{20}/Bi_2Te_3$(001) and $Co_{40}Fe_{40}B_{20}/Bi_2Se_3$(001) heterostructures (further - CoFeB), and studied their crystalline and magnetic properties. It is known [4, 5] that (0001) surface of $V_2VI_3$ monocrystals demonstrates TI properties. Additionally, we have studied crystalline structure and XMCD properties of the CoFeB/BiTeI and $Co_{55}Fe_{45}$/BiTeI (further - CoFe) nanostructures. In [6] TI properties of BiTeI appeared after annealing at 200-250°C were described.

## 1. Experimental

CoFeB/$Bi_2Te_3$(001) and CoFeB/$Bi_2Se_3$(001) layers were grown with use of laser MBE system (produced by Surface, GmbH.) based on KrF excimer laser. Clear surface of TI substrates was obtained using adhesive tape to eliminate top layer of the material. To dehydrate the surface, TI substrates were annealed under $10^{-8}$ mBar pressure at 200°C during 30 minutes. CoFeB layers were grown on $Bi_2Te_3$ at 400°C in two ways: in argon atmosphere under $2.5\times10^{-3}$ mBar pressure and without argon. The thickness of the layers was 10 nm. The CoFeB layer was grown on the $Bi_2Se_3$ substrate in two ways: at RT directly, and on high temperature (400°C) 3-8 Å thick seeding layer with subsequent RT growth of 9-10 nm more. CoFeB/BiTeI nanostructures were grown at RT-350°C temperature range, using conventional UHV MBE system equipped with electron beam metal evaporator. RHEED analysis was performed *in-situ* with use of built-in diffractometer at 30 kV. We have carried out an image analysis via special software developed in our group [7]. The software allows one to plot 3D projections of RHEED patterns in the reciprocal space for concrete zone axis. AFM measurements were realized with the microscope produced by NT-MDT (Zelenograd, Russia).

In order to study the static magnetization response of these CoFeB films, magnetization hysteresis loops were recorded employing angle dependent MOKE in longitudinal geometry measured with a 625 nm laser-diode system at RT at 45° incidence angle geometry, and the change in reflectivity was registered. The hysteresis loops were recorded at different azimuthal angles in the range of 0–360°.

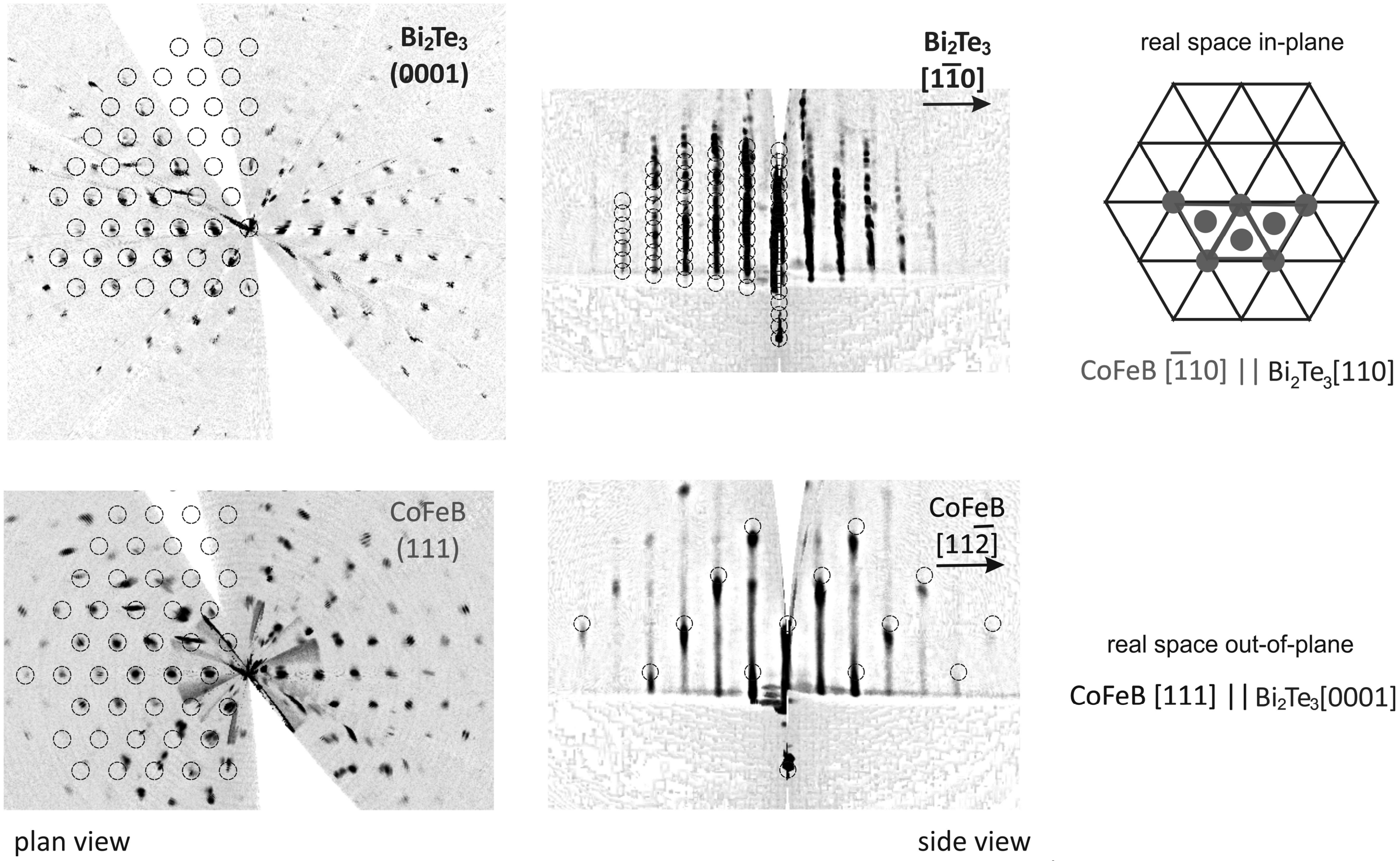

Fig. 1. RHEED reciprocal space maps of CoFeB(111) layer and $Bi_2Te_3$ (001) substrate with superimposed model reflections. RHEED maps from the surface region allow separate study of the substrate and the layer. Schematic presentation of CoFeB(111) / $Bi_2Te_3$(001) in-plane lattice matching is shown on the right.

The chemical composition of deposited layers was studied by XPS using SPECS spectrometer with the PHOIBOS-150-MCD-9 hemispherical energy analyzer and X-ray monochromator FOCUS-500 (Al$K_\alpha$ irradiation, hν=1486.74 eV, 200 W). XAS and XMCD spectra were measured using BL14 beamline synchrotron radiation (HiSOR synchrotron, Hiroshima, Japan).

## 2. Crystalline structure of CoFeB on $Bi_2Te_3$ (or BiTeI) and $Bi_2Se_3$

Fig. 1 shows RHEED reciprocal space maps of CoFeB(001) layer and $Bi_2Te_3$ substrate with superimposed model reflections. The maps on the left show the reciprocal space cross-sections containing a single reciprocal space zone – with the [0001] zone axis for $Bi_2Te_3$ and the [111] zone axis for CoFeB. The maps on the right show the plan-view reciprocal space projections onto the plane parallel to the surface normal. The observed RHEED patterns show pronounced streaks combined with transmission spots in agreement with the moderate surface smoothness observed by AFM. The results of reciprocal lattice modeling are superimposed on the maps in Fig. 1 allowing the conclusion that CoFeB grows epitaxially with the body centered cubic (bcc) crystal structure oriented with its [111] axis perpendicular to the surface. The $a$≈2.84 Å lattice constant of CoFeB is very close to that of $Co_{75}Fe_{25}$ [8]. The in-plane CoFeB [11$\bar{2}$] axis is oriented along the $Bi_2Te_3$ [1$\bar{1}$0] axis favored by the fact that interatomic distance 4.395 Å of $Bi_2Te_3$ in [110] direction is close to the distance 4.019 Å in [$\bar{1}$10] direction of CoFeB.

In case of the CoFeB growth on $Bi_2Se_3$, direct growth of more or less thick layer of the material at elevated temperatures is impossible because of high chemical activity of Se. Nevertheless, 2-3 monolayers thick CoFeB layer demonstrates a set of streaks on a RHEED pattern (Fig. 2 (a, b)) showing an ordering with two times increased period of the structure. Further growth of CoFeB at RT allowed obtaining polycrystalline CoFeB layer (the circles on a RHEED patterns, see Fig. 2 (c)).

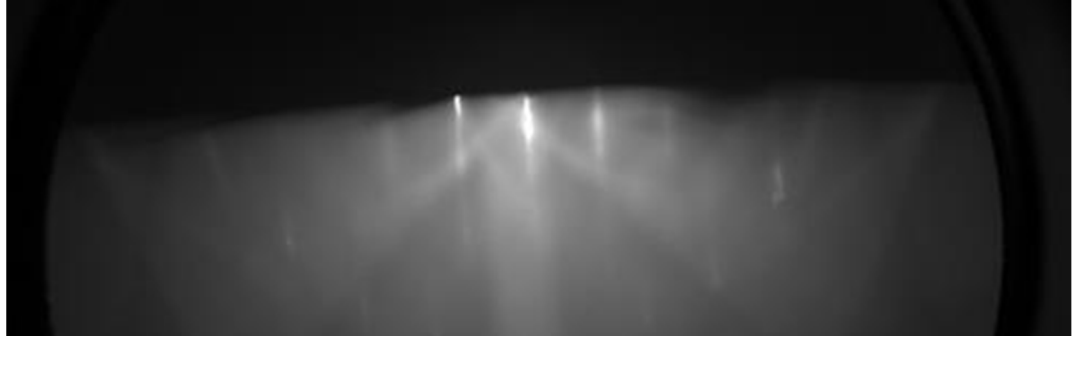

(a)

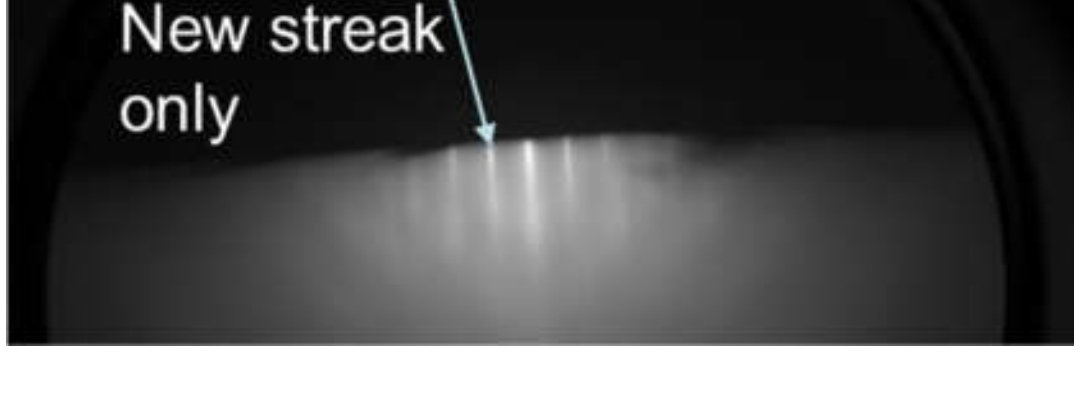

(b)

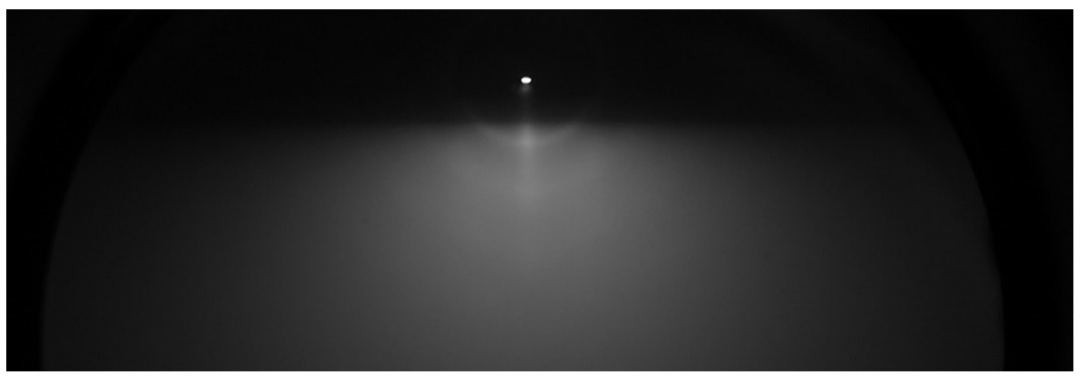

(c)

Fig. 2. Evolution of RHEED pattern during CoFeB growth. (a) – $Bi_2Se_3$ substrate, (b) – about 7 Å of CoFeB grown at 400°C, (c) – about 70 Å of CoFeB grown at RT.

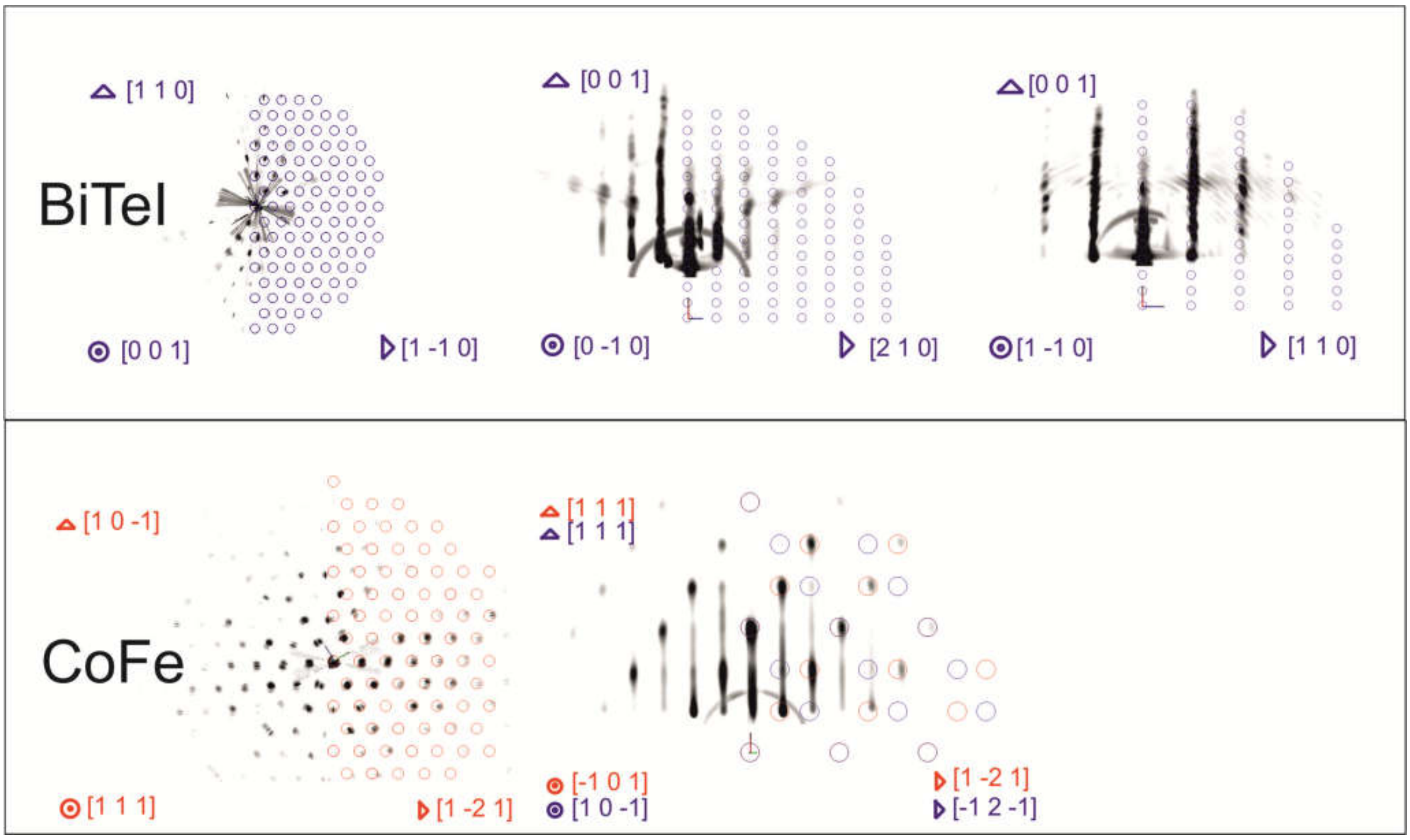

Fig. 3. RHEED reciprocal space maps of CoFe (111) layer and BiTeI (001) substrate with superimposed model reflections. Two sets of axis directions corresponding to two types of structural domains rotated at 180° around [111] axis are shown on the lower right part of the figure.

It was found, that being grown on BiTeI, CoFe and CoFeB also forms bcc-type crystalline structure with the epitaxial relations of the main crystallographic axes slightly differing, as in case of CoFeB/$Bi_2Te_3$ growth (see Fig. 3 for CoFe). Plane (111) of CoFeB is parallel to (001) plane of BiTeI, at that [$10\bar{1}$] (or [$\bar{1}01$]) axis of CoFe coincides with [$0\bar{1}0$] axis of BiTeI.

## 3. MOKE and AFM measurements

Fig.4 (a) and (b) illustrate the azimuthal-angle--dependent ($\phi$) MOKE hysteresis loops of the CoFeB thin films on $Bi_2Te_3$ and $Bi_2Se_3$ substrates, respectively. The MOKE measurements were performed after three months of the sample air exposure. The azimuthal angle was rotated 180° relative to the in-plane magnetic field. The $\phi$-dependent MOKE hysteresis loops of the epitaxial CoFeB alloy film on $Bi_2Te_3$ clearly showed an invariant magnetic coercivity, saturation Kerr signal, and the same shape of hysteresis loop, indicating usual for $6^{th}$-fold symmetry of the substrate an isotropic magnetic behavior with $\phi$ when the magnetic easy axis was on the surface plane. AFM measurements confirm isotropic character of the film (see coalesced triangular islands in Fig. 4(a)). The $\phi$-dependent MOKE hysteresis loops of the CoFeB/$Bi_2Se_3$ clearly exhibited the uniaxial magnetic anisotropy (UMA). As the $\phi$-angle increased from 90 to 0°, the remanence and Kerr signal continuously decreased.

In contrast to the square-shaped and opened hysteresis loop at $\phi$=90°, only a curve with almost zero remanence was observed at $\phi\approx0°$. Above $\phi\approx0°$, the remanence and Kerr signal continuously increased again up to 180°, demonstrated in the inset in Fig. 4(b). This indicates that $\phi\approx0°$ was the magnetic hard-axis direction, whereas $\phi\approx90°$ and 270° were the magnetic easy-axis direction. This observed UMA originated from the shape magnetic anisotropy (see the surface morphology measured by AFM in Fig. 5 (b), in contrast to more isotropic triangular-shaped islands in case of CoFeB/$Bi_2Te_3$, Fig. 5(a)) of $2^{nd}$ order, engendered by microstructure patterning and is actually similar to the UMA observed in other magnetic systems. The nature of such type of shape anisotropy may be related to the presence of the steps on the substrate, like that seen in

Co/$CaF_2$/Si(111) heterostructures [7].

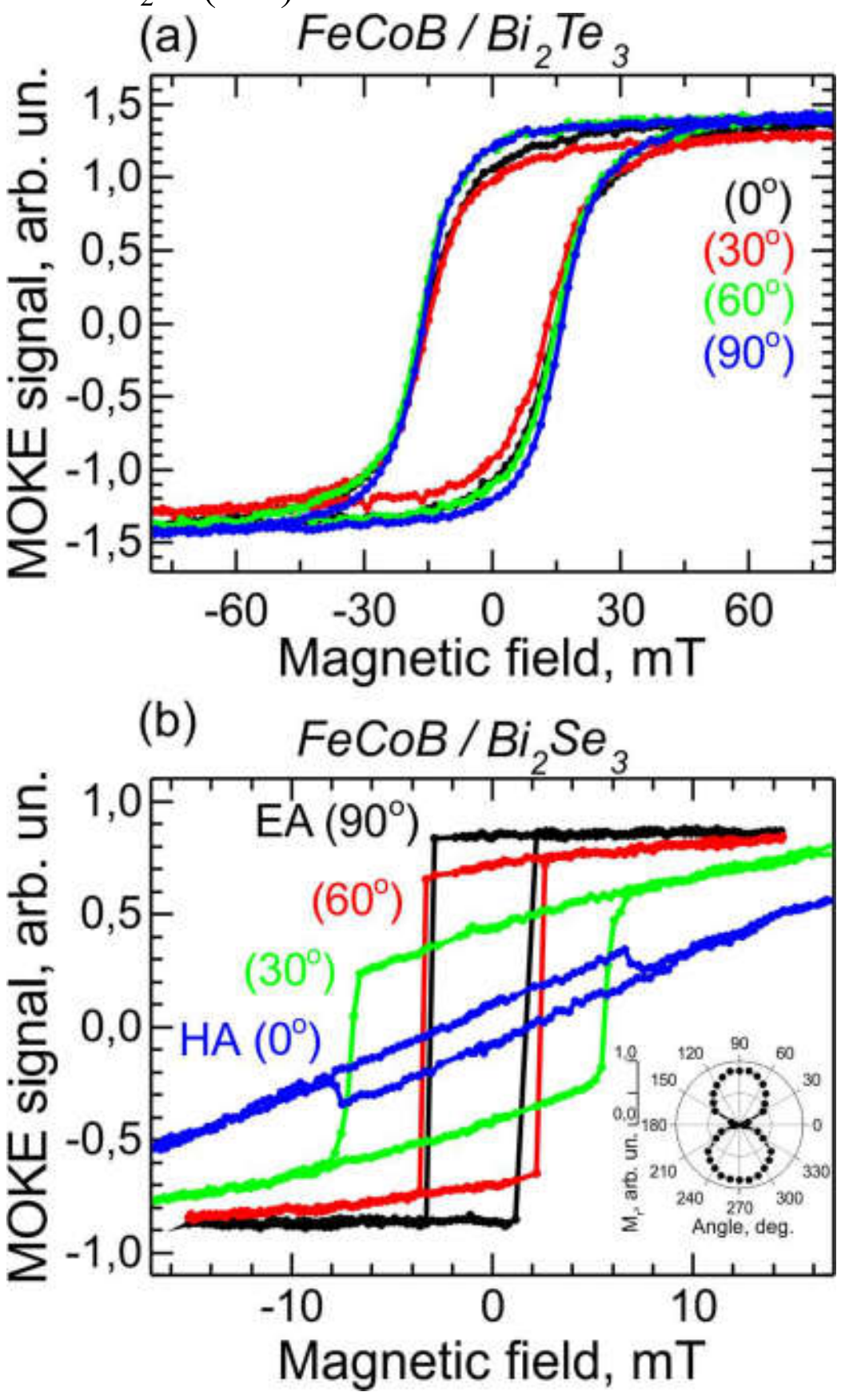

Fig.4. Azimuthal-angle-dependent MOKE hysteresis loops of CoFeB thin films on (a) $Bi_2Te_3$ and (b) $Bi_2Se_3$ substrates.

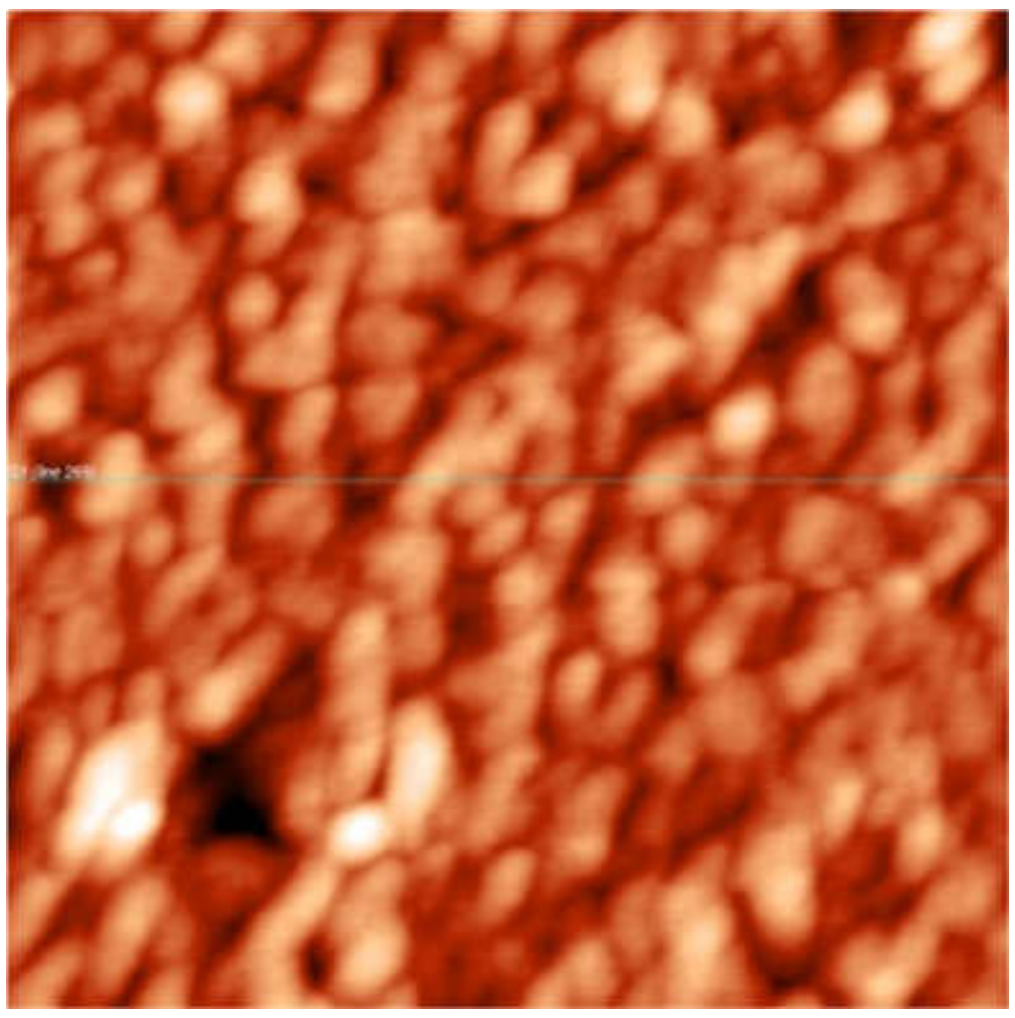

(a)

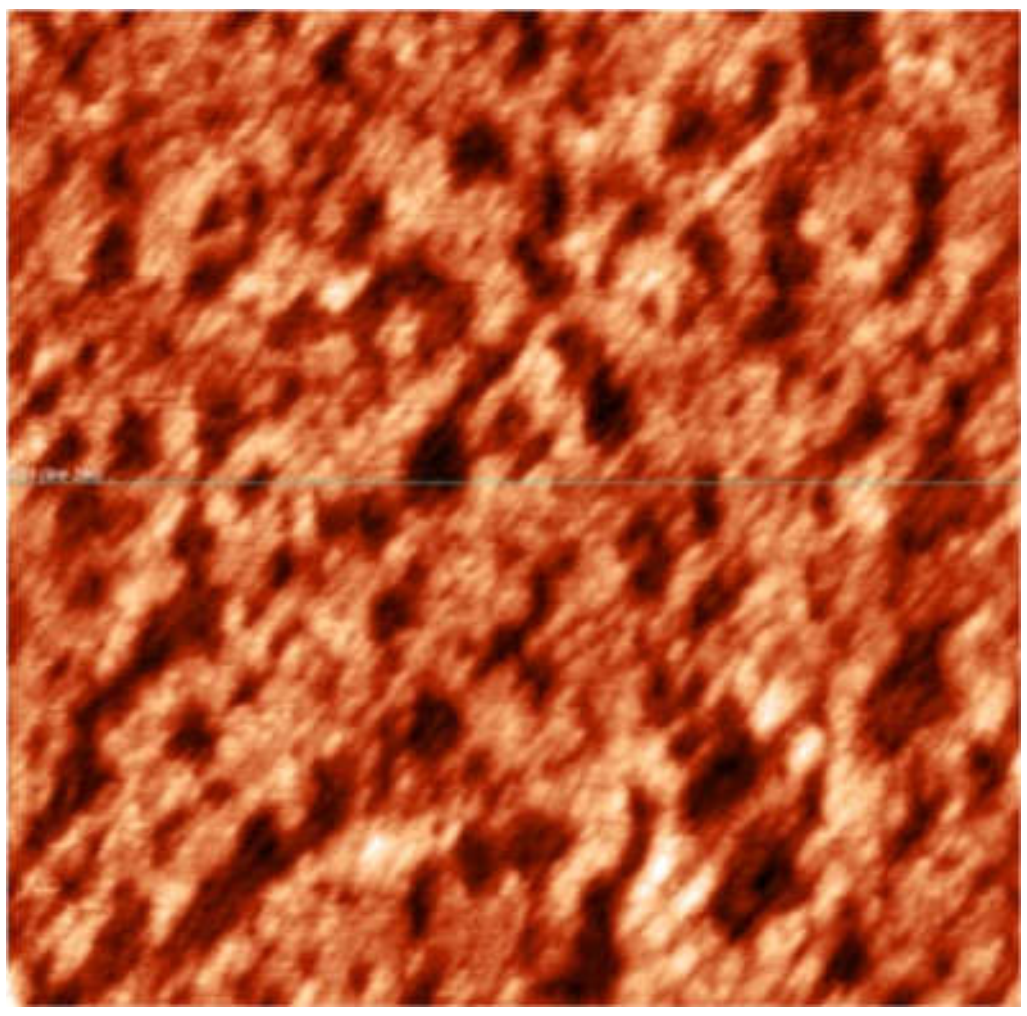

(b)

Fig. 5. AFM measurement results for CoFeB grown on $Bi_2Te_3$ (a, 1 × 1 micrometer$^2$, height 10 nm) and $Bi_2Se_3$ (b, 1 × 1 micrometer$^2$, height 3 nm).

## 4. XAS and XMCD measurements of CoFeB on $Bi_2Se_3$ and CoFeB on $Bi_2Te_3$ (or BiTeI) heterostructures

XAS and XMCD measurements were carried out for Co 2p and Fe 2p edges. The samples were measured at circular polarized X-ray falling beam in appliance of ≤ 0.3 T magnetic field in two opposite directions subsequently. The beam incidence direction is parallel to that of magnetic field. To calculate magnetic moments, the measurements were carried out with the set of the different angles between magnetic field and the sample normal, lying in 25-70° range. Fig. 6 shows Co 2p (a) and Fe 2p (b) $L_{3,2}$ multiplet structure of 5 nm thick CoFeB/$Bi_2Se_3$ grown at RT. The shapes of the spectra correspond to the metallic Co state in these layers [9]. Fe state is almost metallic with small addition corresponding to oxidized Fe (FeO - type) [10].

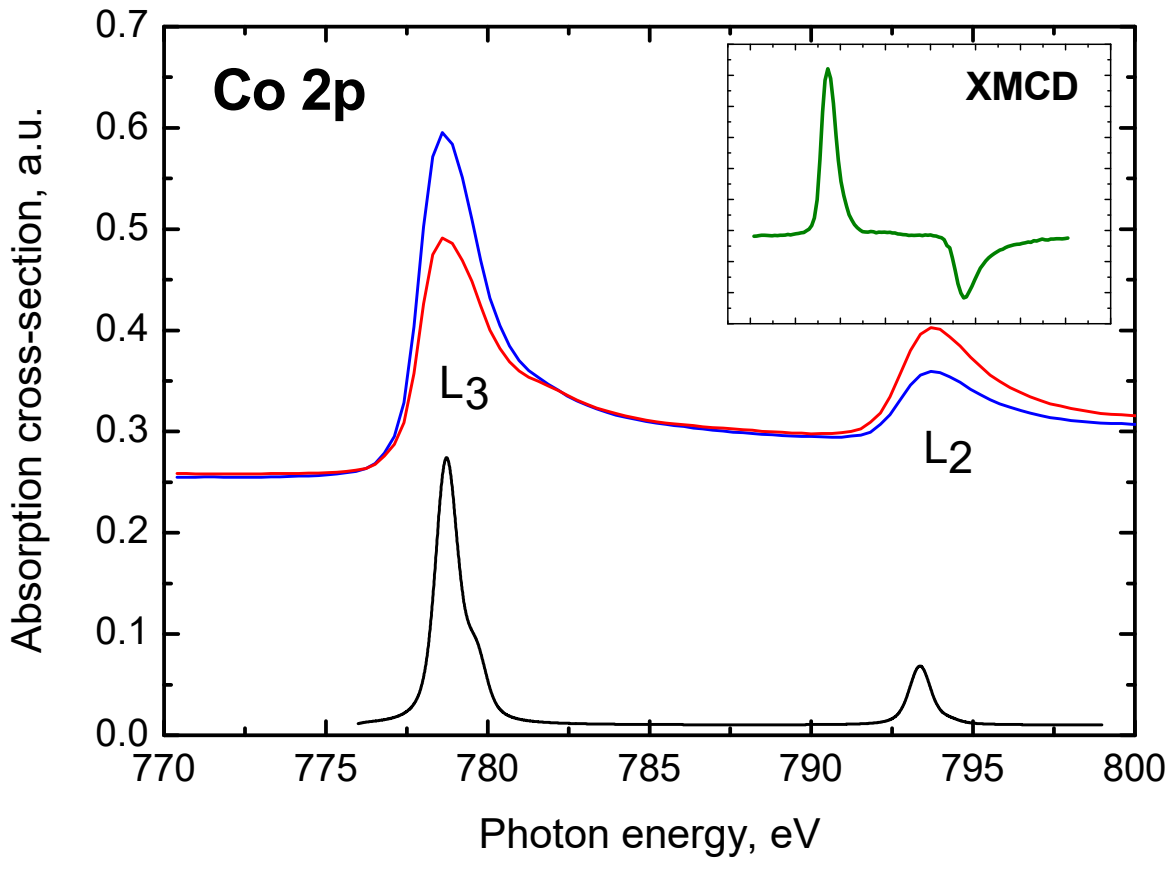

(a)

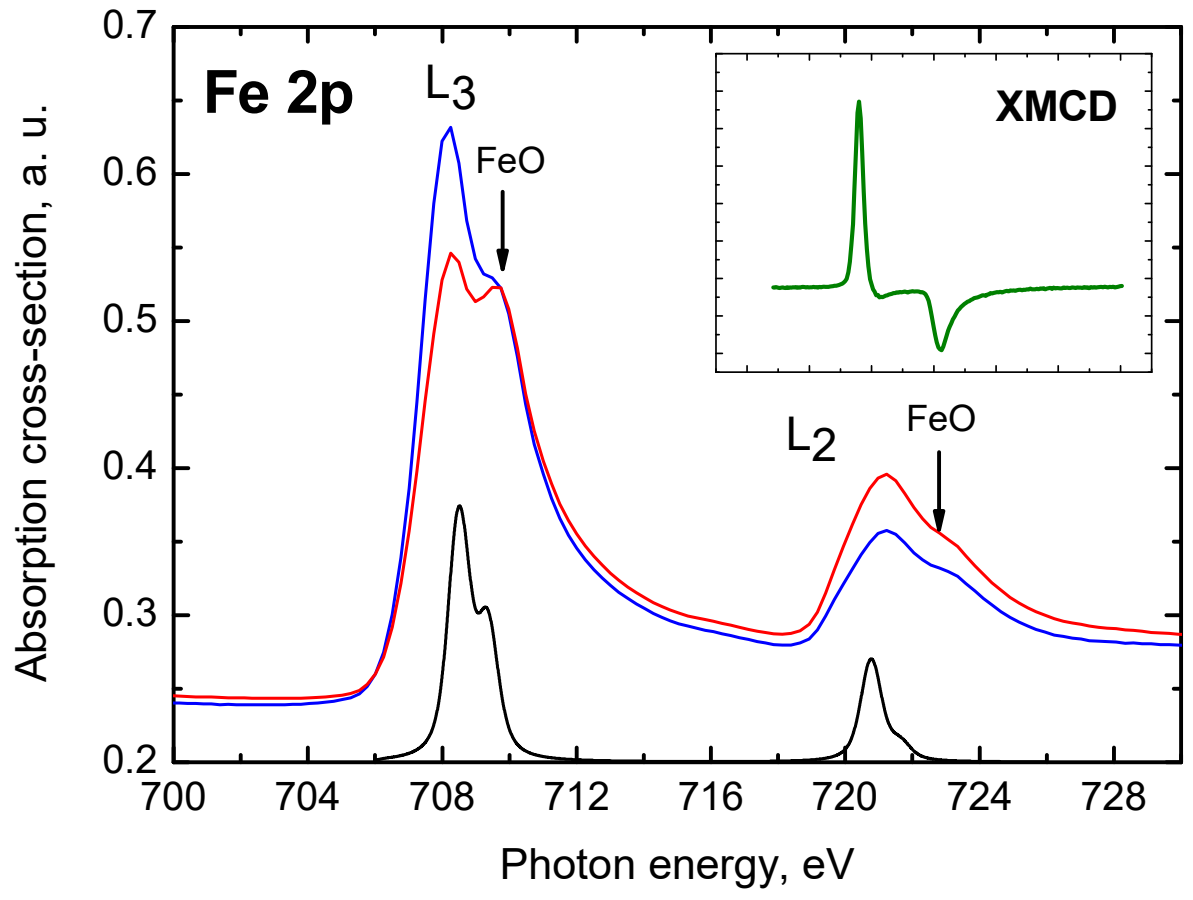

(b)

Fig. 6. Co 2p (a) and Fe 2p (b) $L_{3,2}$ multiplet structures of CoFeB/$Bi_2Se_3$ for two photon circular polarization directions. Bottom lines show simulation results. XMCD signals are shown in the insets.

The shapes of the spectra were simulated with use of code CTM4XAS ([11]). The results of the simulation are shown with the bottom curves on Fig. 6. In $O_h$ coordination 3d state is split to $t_{2g}$ and $e_g$ ($\Delta E$=10Dq). Simulation was done for $Co^{2+}$ and $Fe^{2+}$ with 10Dq=0.2 eV. Slater integral reduction $F_{dd}$ = 0.4, $F_{pd}$ = $G_{pd}$ = 0.

Measured XMCD signals allow the calculation of spin and orbital magnetic moments for Fe and Co in the film. Magnetic moments (both in-plane and out-of-plane projections) were calculated using original software developed by the authors of this work.

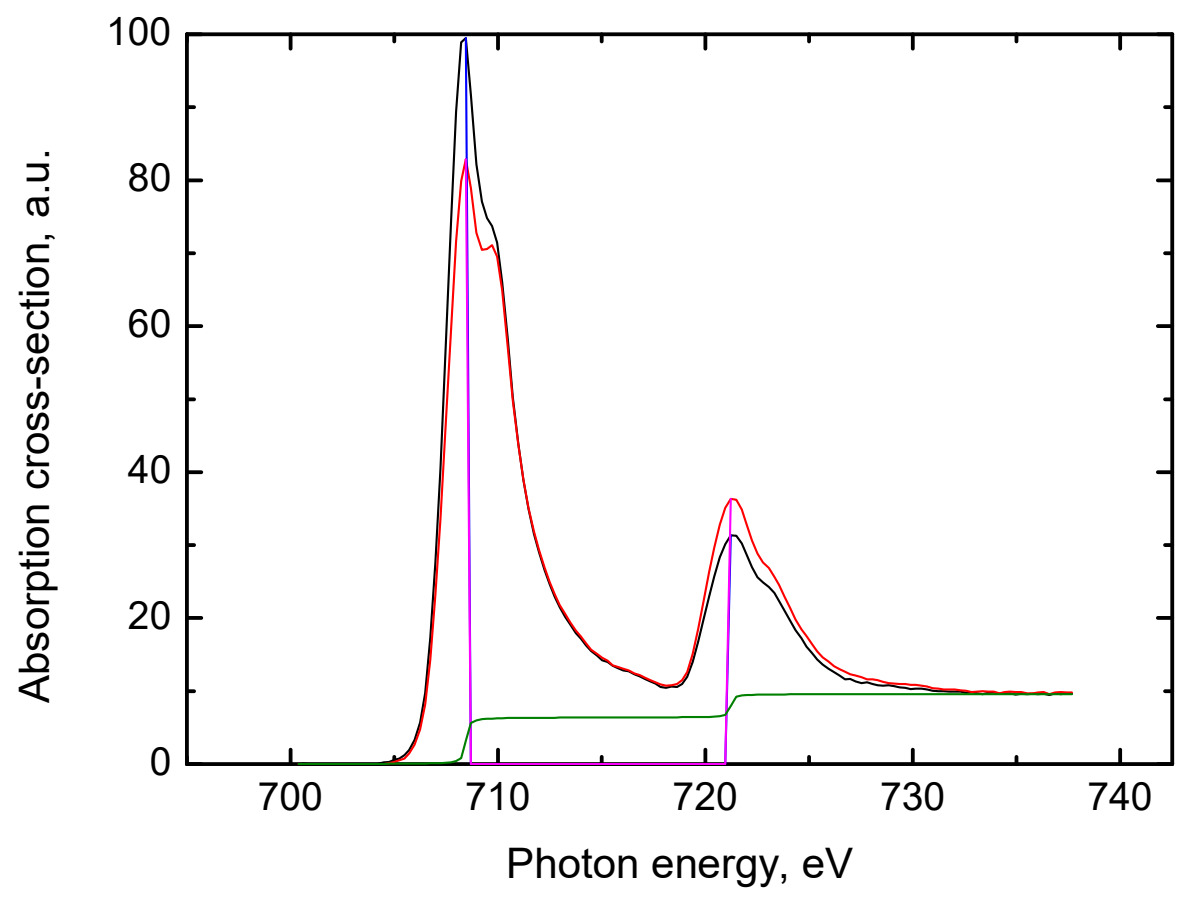

Fig. 7. (Colored online) Explanation of the calculation steps of spin and orbital magnetic moments. Black and red – normalized XAS spectra for two photon polarization (or two external magnetic field opposite directions) after linear background subtraction, green – stepped background being subtracted, magenta – pointing the maxima for the calculations.

Common principle of the spin and orbital moment calculation of 3d elements according to the sum rules was shown in [12, 13]. The measurements were carried out in two geometries. At first, the sample is situated perpendicular to the photon direction. At second, the sample is situated at the intermediate angle (between 0 and 90°) to the photon direction. Instead of change the photon circular polarization direction, one may change an applied magnetic field direction to the opposite one.

Usual way to calculate magnetic moments contain seven steps:

1) Normalization of pair measured XMCD curves;
2) Subtraction the linear background;
3) Subtraction the stepped background (see Fig. 7);
4) Calculation of the sum areas under peaks for each curve;
5) Calculation of the areas $\Delta A_{L3}$ and $\Delta A_{L2}$ of XMCD curve peaks.
6) Calculation of experimental effective spin and orbital moments according to the formula [14]

$$m_s^{eff}(\theta) = -2\left(\frac{N_h\mu_B}{P_c}\right)\frac{\Delta A_{L3}-\Delta A_{L2}}{A_{L3}+A_{L2}},$$
$$m_{orb}(\theta) = -\frac{4}{3}\left(\frac{N_h\mu_B}{P_c}\right)\frac{\Delta A_{L3}+\Delta A_{L2}}{A_{L3}+A_{L2}} \quad (1),$$

where $N_h$ – population of the studied element $d$ – level, $P_c$ – photon circular polarization degree, $A_{L3}$ and $A_{L2}$ – sum areas under appropriate $L_3$ and $L_2$ maxima for pair of XAS curves.

7) Solution of the system of four linear equations, where the first pair of equations is [15]

$$m_s^{eff}(\theta) = m_s\cos(\theta-\varphi) + 7m_T^{\perp}\cos\theta\cos\varphi + 7\,m_T^{\parallel}\sin\theta\sin\varphi,$$

$$m_{orb}(\theta) = m_{orb}^{\perp}\cos\theta\cos\varphi + m_{orb}^{\parallel}\sin\theta\sin\varphi \quad (2),$$

the second one is the same for another geometry (i.e., another pair of $\theta$ and $\varphi$ values, see Fig. 8(a)). Here the values of the sought-for spin and orbital magnetic moments $m_s$ and $m_{orb}$ may be found. The value of $m_T$ is the product of a second rank tensor, the quadrupole moment, and the spin moment [12]. Most popular special case is the case where $\theta_1$ = 0 (normal incidence for perpendicular magnetization, Fig. 8(b)), and $\theta_2$ = 30-60° (grazing incidence for in-plane magnetization, Fig. 8(c)). The system for this case will be

$$m_s^{eff}(0) = m_s + 7m_T^{\perp},$$
$$m_{orb}(0) = m_{orb}^{\perp},$$
$$m_s^{eff}(\theta_2) = (m_s - 3.5m_T^{\perp})\sin\theta_2,$$
$$m_{orb}(\theta_2) = m_{orb}^{\parallel}\sin\theta_2 \quad (3).$$

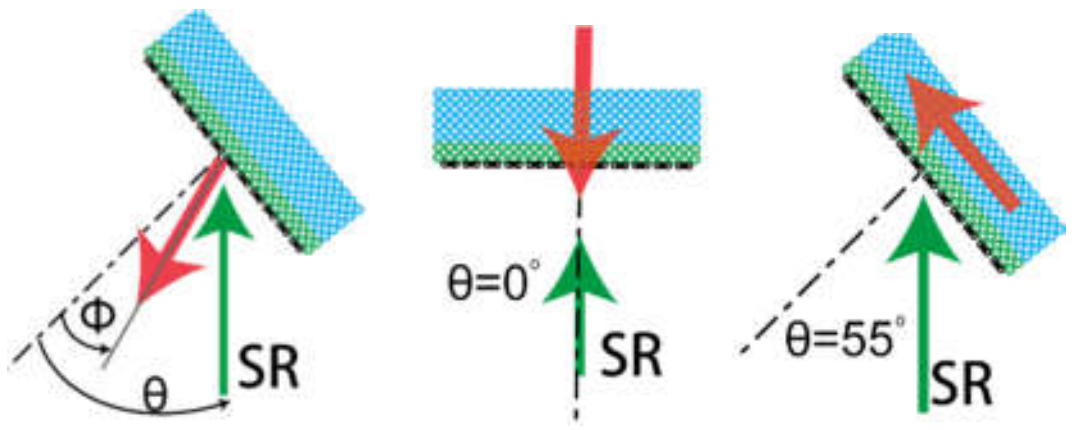

(a) (b) (c)

Fig. 8. (a) - geometry of the XMCD experiment. (b) - normal incidence for perpendicular magnetization, (c) - grazing incidence for in-plane magnetization, for example 55° (an external magnetic field will rotate magnetization vector out of the sample plane). Here $\theta$ is the angle between photon wave vector and surface normal, φ - angle between magnetization direction and surface normal.

In the experiments carried out we supposed that at all falling angles the saturation occurs, i.e. magnetic field is strong enough to rotate magnetization vector along the beam direction, i .e. $\varphi = \theta$ in Fig. 8 (a). The confirmation of this supposition is the behavior of element-dependent magnetic hysteresis loops measured using XAS method. In this method, TEY current is measured from the sample with the applied external magnetic field tuned in -0.3 T to 0.3 T range, at two values of fixed X-ray energy: at the $L_3$ peak maximum and out of it.

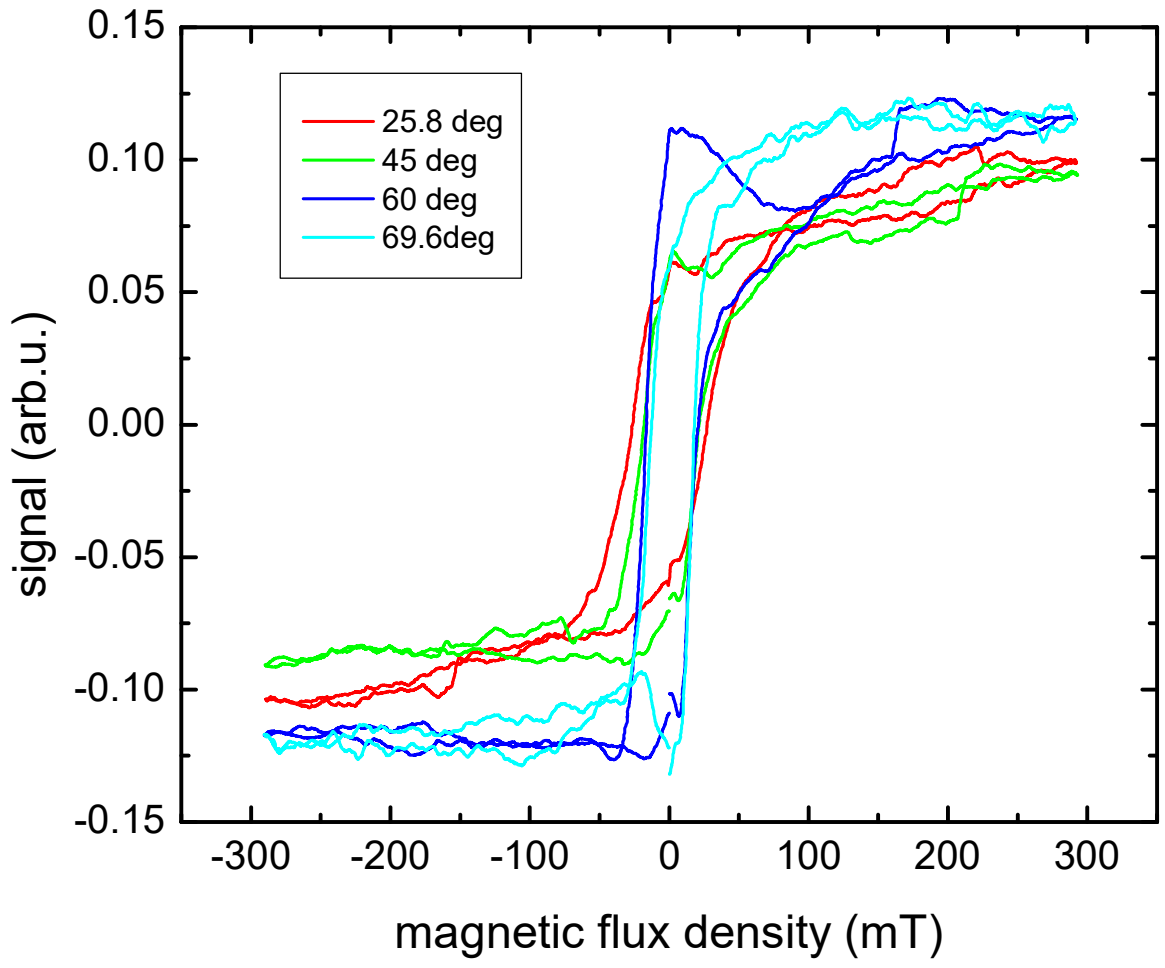

Fig. 9. The set of hysteresis loops measured by XAS for Fe $L_3$ edge with different values of the photon beam incident angle.

In Fig. 9 the set of loops measured for Fe $L_3$ edge at the different incident angles $\theta$ for Fe is shown. It is seen that at the 0.3 T external magnetic field (used in the measurements) all the loops are saturated. The behavior of coercivity and remanence on the $\theta$ angle is characteristic for easy plane – type magnetic anisotropy.

The results of the calculations for 5 nm thick CoFeB grown on $Bi_2Se_3$ at RT are shown in the upper part of Table 1.

| CoFeB/$Bi_2Se_3$ | | |
|---|---|---|
| | Fe | Co |
| $m_s$ calc. (bulk [16]) | 1.12(1.98) | 1.64 (1.55) |
| $m_{orb}$ calc. (bulk [16]) | 0.16 (0.06-0.08) | 0.38 (0.15) |
| $m_{orb\,\parallel}$ calc. | 0.14 | 0.33 |
| $m_{orb\,\perp}$ calc. | 0.08 | 0.19 |
| CoFeB/$Bi_2Te_3$ | | |
| | Fe | Co |
| $m_s$ calc. | 1.62 | 1.23 |
| $m_{orb}$ calc. | 0.1 | 0.161 |
| $m_{orb\,\parallel}$ calc. | 0.03 | 0.009 |
| $m_{orb\,\perp}$ calc. | 0.09 | 0.16 |
| CoFeB/BiTeI | | |
| | Fe | Co |
| $m_s$ calc. | 1.92 | 1.82 |
| $m_{orb}$ calc. | 0.17 | 0.21 |
| $m_{orb\,\parallel}$ calc. | 0.05 | 0.04 |
| $m_{orb\,\perp}$ calc. | 0.16 | 0.20 |

Table 1. Calculated values of spin and orbital moments (including in-plane and out-of-plane components of $m_{orb}$) of Co and Fe in CoFeB in compare to bulk values (note that the last ones vary in different papers) for CoFeB/$Bi_2Se_3$ and CoFeB/$Bi_2Te_3$.

The results of XAS and XMCD measurements of CoFeB/$Bi_2Te_3$ heterostructures are shown in Fig. 10.

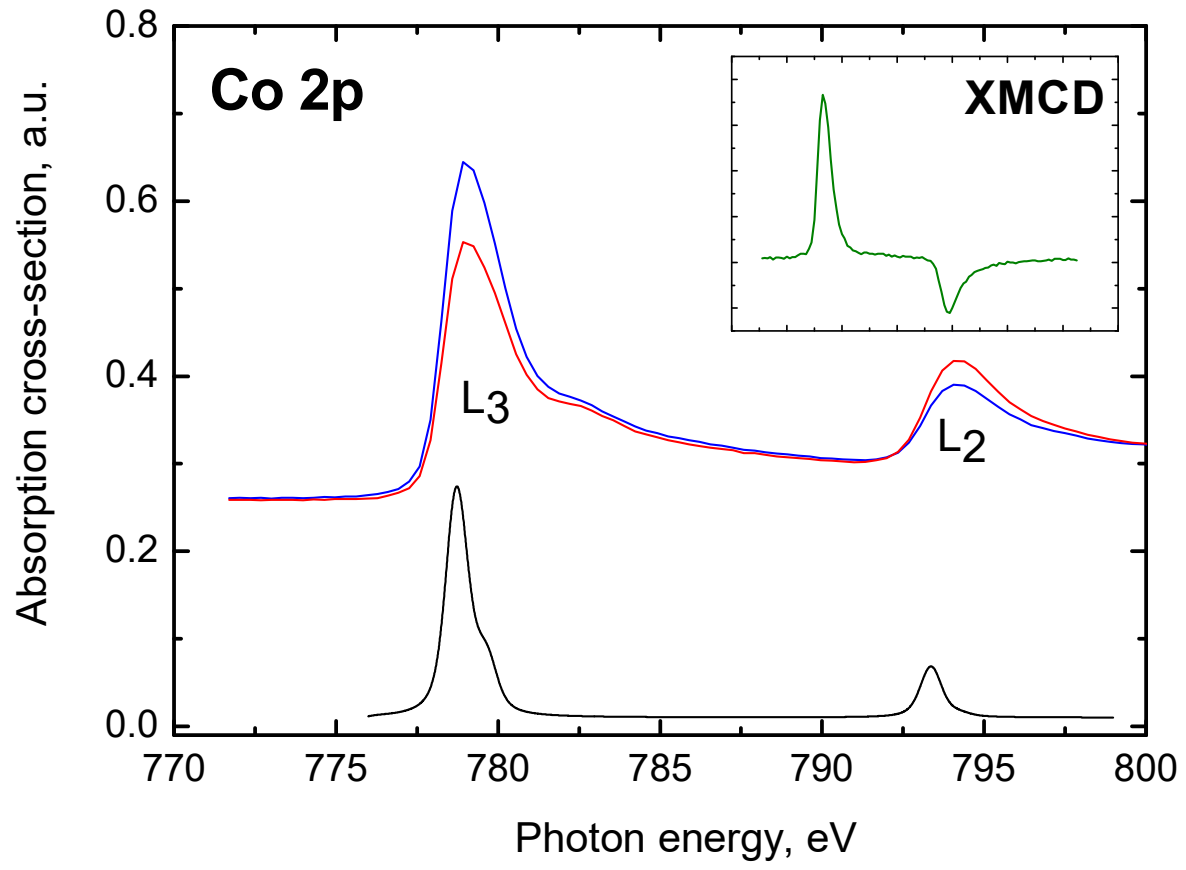

(a)

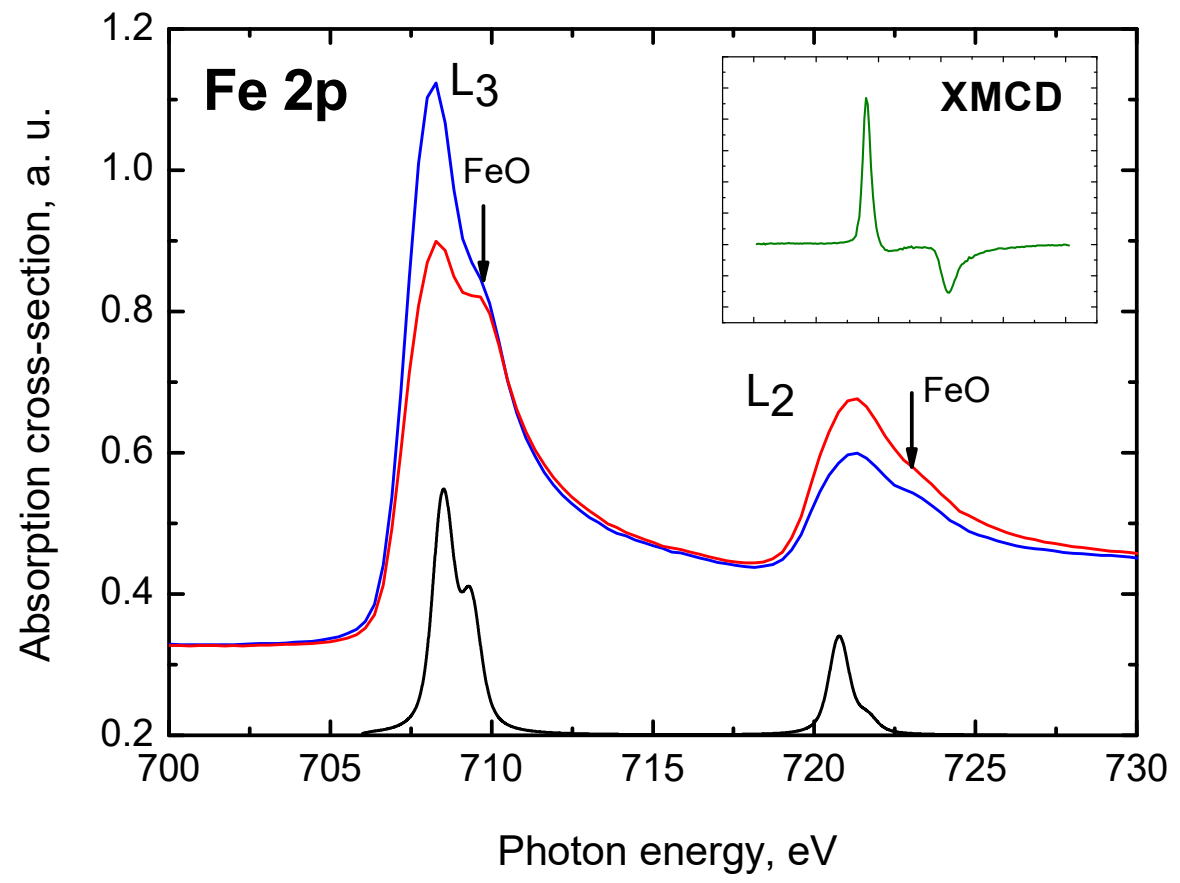

(b)

Fig. 10. Co 2p (a) and Fe 2p (b) $L_{3,2}$ multiplet structures of $CoFeB/Bi_2Te_3$ for two photon circular polarization directions. Bottom lines show simulation results. XMCD signals are shown in the insets.

As in case of $CoFeB/Bi_2Se_3$ described above, here in $O_h$ coordination 3d state is split to $t_{2g}$ and $e_g$ ($\Delta E$=10Dq). Simulation was done for $Co^{2+}$ and $Fe^{2+}$ with 10Dq=0.2 eV. Slater integral reduction slightly differs from the $CoFeB/Bi_2Se_3$ case: $F_{dd}$ = 0.38, $F_{pd}$ = $G_{pd}$ = 0. The results of the calculations for 5 nm thick CoFeB grown on $Bi_2Te_3$ at 300°C are shown in the middle part of Table 1.

Table 1 shows that the values of $m_s$ are smaller than that for bulk metallic Fe for both $CoFeB/Bi_2Se_3$ and $CoFeB/Bi_2Te_3$ (especially in the first case). The value of $m_s$ for Co is close to that for the metallic state for the case of $CoFeB/Bi_2Se_3$ and slightly smaller for the case of $CoFeB/Bi_2Te_3$. The values of $m_{orb}$ are slightly larger than that for the metallic state in case of $CoFeB/Bi_2Se_3$, and close to that for the metallic state in case of $CoFeB/Bi_2Te_3$. Some authors [17-19] state that decrease of $m_s$ may be related to the decrease of the system dimension (thin layers, monoatomic chains etc.) or to the presence of the oxidized state of metal.

In case of CoFeB and CoFe growth on the BiTeI substrate, the shape of the XAS spectra and XMCD signals is very similar to that for $CoFeB/Bi_2Te_3$ shown in Fig. 10. The values of spin and orbital magnetic moments are shown in the lower part of Table 1. They are relatively close to that for the metallic Fe and Co states. This closeness is maximal in case of the using of the $LaAlO_3$ capping layer preserving the oxidation at atmosphere.

The works to increase the precision of seven-step procedure described above are underway.

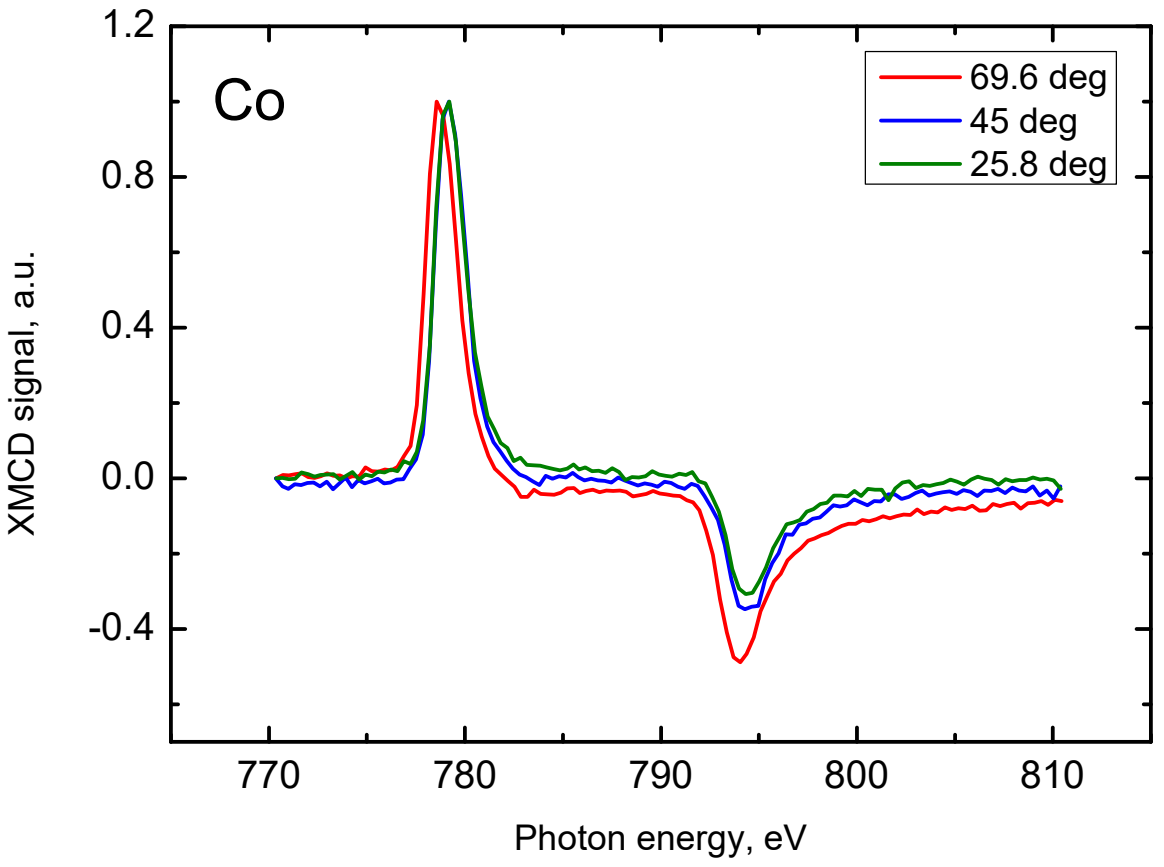

(a)

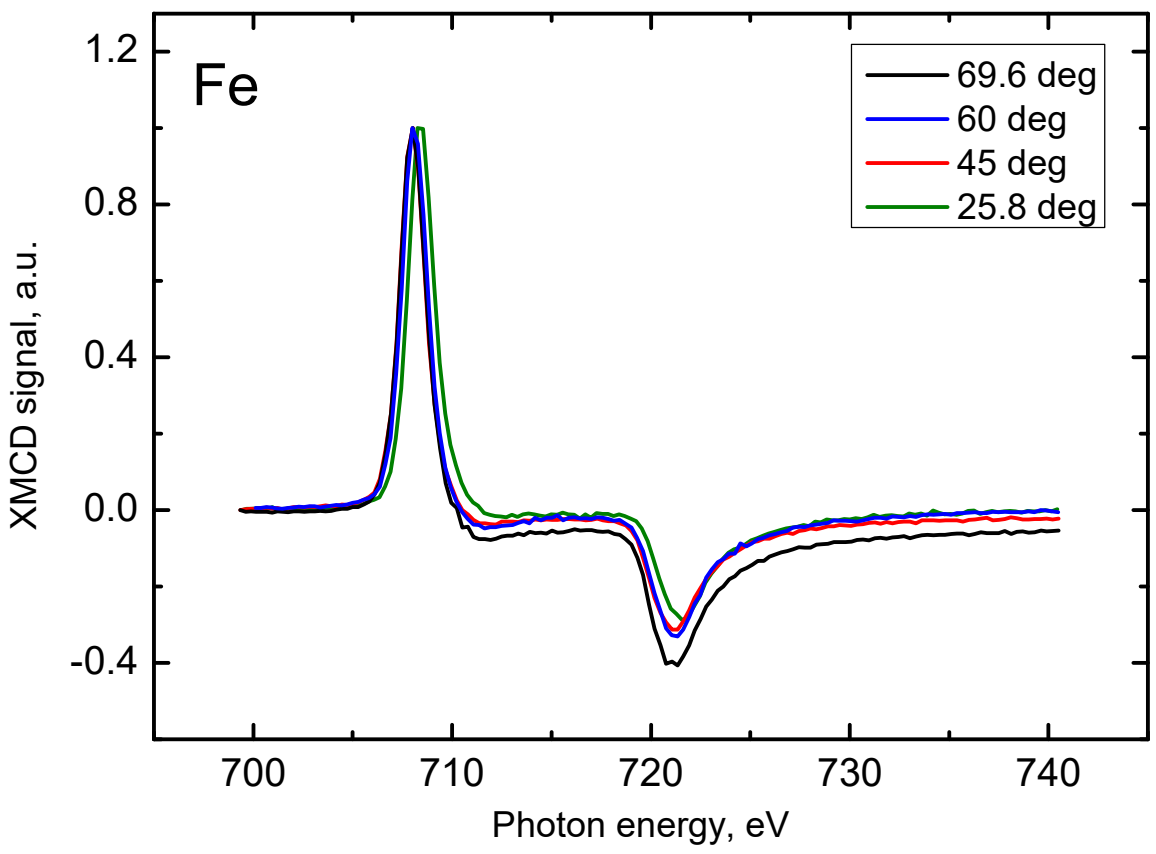

(b)

Fig. 11. The angular dependence of the XMCD signal of Co (a) and Fe (b) in $CoFeB/Bi_2Te_3$ nanostructure.

For the case of $CoFeB/Bi_2Te_3$ the angular dependence of the XMCD signal is measured. In Fig. 11 the dependence of the XMCD signals normalized to $L_3$ peak height on the photon energy for different beam incident angles $\theta$ is shown for Co (a) and Fe (b). It is seen that the value of $\Delta A_{L2}$ decreases with $\theta$ value decrease. This fact confirms that the values of calculated experimental effective spin and orbital moments (1) definitely depend on the XMCD signal shape (i.e. depend on θ value).

## 5. XPS measurements

The XPS investigation was carried out to explore the evolution of the elemental chemical states at the CoFeB/TI surface during air exposing. Fig. 12 shows the portions of XPS spectra corresponding to the Co-$2p_{3/2}$, Fe-$2p_{3/2}$ and B1s peaks recorded on the CoFeB/ $Bi_2Te_3$ (top row) and $CoFeB/Bi_2Se_3$ (bottom row) structures for as-deposited (black curves) and Ar-sputtered samples (red curves). For the $CoFeB/Bi_2Te_3$ structure, Co 2p peak consists of several peaks and according to the XPS handbook [20], the peak located at 778.3 eV corresponds to the metallic Co2p3/2 line and the peak at 780.5 eV is related to the CoO. For sputtered sample (red curves), the peak position is not shifted compared with as-deposited one. However, CoO intensity shows a dramatic decrease due to removing of top oxides layer. A wide Fe $2p_{3/2}$ peaks (Fig. 12 (b)) can be identified as Fe 2p3/2 peak in metallic Fe (706.7 eV), Fe $2p_{3/2}$ peak in $FeO_x$ (710.0 eV), and Auger peak from Co (713.0 eV) (shown by arrow). For sputtered sample (red curves), $FeO_x$ intensity decreases but not so strong like in the case of Co peak, meaning more deeper distribution of $FeO_x$.

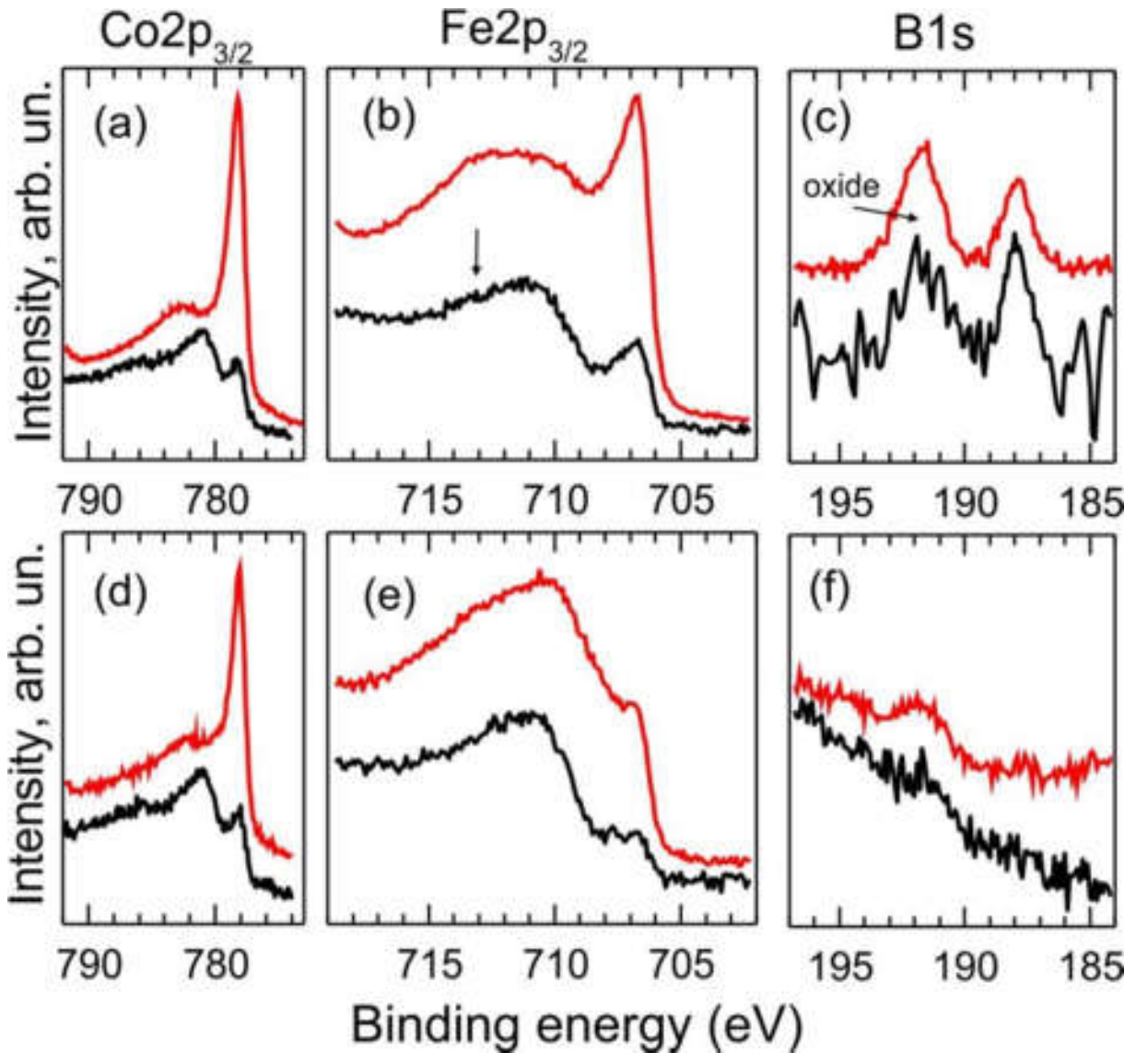

Fig. 12. XPS spectra corresponding to Co 2p (a, d), Fe 2p (b. e) and B 1s (c, f) electronic levels recorded on CoFeB 5 nm thick film deposited on $Bi_2Te_3$ (top row) and $Bi_2Se_3$ (bottom row), respectively. Black curves correspond to the as-deposited samples exposed in the air for longer than 3 months. Red curves were measured after 2 minutes of Ar+ etching corresponding of 0.5 nm top layer.

In the B 1s signal from as-deposited surface and after sputtering, two components can be distinguished. Because the B 1s signal is very weak, we need to increase a lot the scan number and dwell time to increase the signal to noise ratio. The lower binding energy one peak at about 188 eV is assigned to the metallic boron while the other peak at about 193 eV can be assigned to the oxidized boron. One can see that after removing of about 0.5 nm oxidized top layer the ratio between metallic and oxidized components kept almost the same.

For the CoFeB/$Bi_2Se_3$ structure, Co 2p peak behaves in a similar way to that observed on the CoFeB/ $Bi_2Te_3$ (Fig. 12(d)), while Fe and B are mostly in the oxidized states even after removing of 0.5-1.0 nm of top layer. Thus, the XPS results show a distinct difference in chemical states in cases of CoFeB growth on $Bi_2Se_3$ and $Bi_2Te_3$ substrates. Epitaxial CoFeB film grown on $Bi_2Te_3$ is more stable to oxidation (which is in agreement with the XMCD measurements data discussed above) and demonstrates an isotropic in-plane magnetic behavior. Note, that on both metal covered TI surfaces the Bi and Te(Se) were detected in the amount of about percentage concentrations caused by the effect of segregation and interdiffusion in agreement with recent theoretical calculation.

## Conclusion

In conclusion, we have obtained and studied structurally ordered ferromagnetic films on the topological insulator surface for the first time. Bcc-type CoFeB structural modification is formed on $Bi_2Te_3$(0001), during the growth at 200-400°C. Epitaxial relations of main crystallographic axes of the substrate and grown layer were found, (111) plane of CoFeB is most advantageous as a growth plane, because of $6^{th}$ order of symmetry which is the same as that of the substrate. Polycrystalline CoFeB layer was obtained on the $Bi_2Se_3$ substrates using high-temperature seeding layer. MOKE measurements showed the difference of the azimuthal behavior of magnetization vector in both cases, which may be attributed to the features of the surface morphology revealed by AFM.

XPS measurements demonstrate a distinct difference in chemical states in cases of CoFeB growth on $Bi_2Se_3$ and $Bi_2Te_3$ substrates: being grown on $Bi_2Te_3$, CoFeB is more stable to the oxidation in atmosphere.

XAS and XMCD measurements allowed calculation of the spin and orbital moments of Co and Fe in CoFeB grown on the $Bi_2Se_3$ and $Bi_2Te_3$ films. It was found that the most deviated value of the Fe spin moment from that for bulk metallic Fe occurs in case of more oxidized material presence, i.e. in case of CoFeB/$Bi_2Se_3$ growth, which is in concordance with XPS data.

The results allow us to proceed to the electrical and magnetic measurements of structurally ordered CoFeB/TI heterostructures in appliance to TI-based SpinFET research and development.

*Acknowledgement*

This work has been supported by Russian Foundation of Basic Research (grant № 17-02-00729). The study has also been supported by the Russian Science Foundation (Project No.17-12-01047) in part of the single TIs crystal growth, MOKE and XPS characterization.